\documentclass[amsmath,amssymb,aps]{revtex4}

\usepackage{graphicx}
\begin{document}

\title{Nonextensive statistical effects in the hadron to quark-gluon
phase transition}

\author{A. Lavagno, D. Pigato, P. Quarati}

\affiliation{Dipartimento di Fisica, Politecnico di Torino, C.so
Duca degli Abruzzi 24,  Italy \\ Istituto Nazionale di Fisica
Nucleare (INFN), Sezione di Torino, Italy}

\begin{abstract}
We investigate the relativistic equation of state of hadronic
matter and quark-gluon plasma at finite temperature and baryon
density in the framework of the nonextensive statistical
mechanics, characterized by power-law quantum distributions. We
study the phase transition from hadronic matter to quark-gluon
plasma by requiring the Gibbs conditions on the global
conservation of baryon number and electric charge fraction. We
show that nonextensive statistical effects play a crucial role in
the equation of state and in the formation of mixed phase also for
small deviations from the standard Boltzmann-Gibbs statistics.
\end{abstract}

\maketitle

\section{Introduction}\label{intro}

The physics of high energy heavy ion collisions is a goldmine of
problems in statistical mechanics and thermodynamics due to a large
average number of particles involved and possible phase transition
phenomena in the hot and dense fireball created during the
collisions \cite{hwa}.
%The necessity on statistical physics is especially obvious in the
%study of heavy ion collisions where large average number of
%particles is involved during the collision \cite{hwa}....
In relativistic heavy ion collisions the baryon density can reach
values of a few times the saturation nuclear density and/or high
temperatures. Furthermore, the future CBM (Compressed Baryonic
Matter) experiment of FAIR (Facility of Antiproton and Ion Research)
at GSI Darmstadt, will make possible to create compressed baryonic
matter with a high net baryon density \cite{senger,arsene,bravina}.
In this direction interesting results have been obtained at low SPS
energy and are foreseen at a low-energy scan at RHIC
\cite{afa,alt,hohne,caines,saka}.

Lattice calculations predict a critical phase transition from
hadronic matter to quark-gluon plasma (QGP) at temperature $T_c$
of about 170 MeV, corresponding to a critical energy density
$\epsilon_c\approx$ 1 GeV/fm$^3$ \cite{karsch}. On the other hand,
in dense nuclear matter, baryons are forced to stay so close one
to another that they would overlap. At large densities,
constituent quarks are shared by neighboring baryons and should
eventually become mobile over a distance larger than the typical
size of one baryon. This means that quarks become deconfined and
that at large densities and/or high temperatures they are the real
degrees of freedom of strongly interacting matter instead of
baryons. The process of deconfinement and the EOS of quark-gluon
matter can in principle be described by quantum chromodynamics.
However, in energy density range reached in relativistic heavy-ion
collisions, non-perturbative effects in the complex theory of QCD
are not negligible \cite{karsch}. The generated QGP in the early
stages of the collisions does not at all resemble a quasi-ideal
gas of quarks and gluons because strongly dynamical correlations
are present, including long-range interactions
\cite{hwa,hei,ropke,biro,biro08}. In the absence of a converging
method to approach QCD at finite density one often turns to
(effective) model investigations. Various results from QCD
inspired models indicate that, increasing the baryon chemical
potential in the phase diagram, a region of non-singular but rapid
cross-over of thermodynamic observable around a quasi-critical
temperature, leads to a critical endpoint (CEP), beyond which the
system shows a first order phase transition from confined to
deconfined matter. The existence or exclusion of a CEP has not yet
been confirmed by QCD lattice simulations. Actually, there are
some extrapolation techniques to finite chemical potentials
\cite{schmidt}, although the precise location of the CEP is still
a matter of debate \cite{schaefer}. For example, in Ref.
\cite{fodor}, the authors estimate the values $T^{\rm CEP}=162$
MeV and $\mu^{\rm CEP}=360$ MeV. Such a CEP can be in principle
detected in future high-energy compressed nuclear matter
experiments.

Recently, there is an increasing evidence that the nonextensive
statistical mechanics, proposed by Tsallis, can be considered as
an appropriate basis to deal with physi\-cal phenomena where
strong dynamical correlations, long-range interactions and
microscopic memory effects take place
\cite{tsallis,GMTsallis,book2,kodama}. A considerable variety of
physical applications involve a quantitative agreement between
experimental data and theoretical models based on Tsallis'
thermostatistics. In particular, in the last years there is a
growing interest to high energy physics applications of
nonextensive statistics \cite{epja} and several authors have
outlined the possibility that experimental observations in
relativistic heavy ion collisions can reflect nonextensive
statistical behaviors
\cite{bediaga,albe,beck,rafelski,wilk1,plb2001,biro04,biroprl05,gorenstein,wilk2,physicaA2008,cley,chinellato}.

The existence of nonextensive statistical effects should strongly
affects the finite temperature and density nuclear equation of
state (EOS) \cite{miller,physicaEOS,braz}. In fact, by varying
temperature and density, the EOS reflects in terms of the
macroscopic thermodynamical variables the microscopic interactions
of the different phases of nuclear matter. The extraction of
information about the EOS at different densities and temperatures
by means of heavy ion collisions is a very difficult task and can
be realized only indirectly by comparing the experimental data
with different theoretical models, such as, for example,
fluid-dynamical models \cite{iva}. Related to this aspect, it is
relevant to observe that a relativistic kinetic nonextensive
theory \cite {pla2002} and a nonextensive version of a
hydrodynamic model for multiparticle production processes have
been proposed \cite{wilk08}. Very recently, nonextensive
statistical effects on the hadronic EOS have been investigated by
means of a Walecka type relativistic mean field model
\cite{silva}. Furthermore, a nonextensive version of
Nambu-Jona-Lasinio model \cite{wilknjl} and the effects on color
superconducting phase for two quark flavors due to a change to
Tsallis statistics have been studied \cite{miller2}.
%Preliminary
%studies on the hadron-quark phase transition in the framework of
%nonextensive thermostatistics have been considered in Ref.
%\cite{braz}.

The main goal of this paper is to study how nonextensive
statistical effects influence, from a phenomenological point of
view, the nuclear EOS and, as a consequence, the relative phase
transition at finite temperature and density reachable in
high-energy heavy-ion collisions. Focusing our investigation to
lower temperatures and higher baryon chemical potentials than the
corresponding CEP values, a mixed phase of hadrons, quarks and
gluons can be formed following the Gibbs conditions for the phase
equilibrium. If, in general, we consider a substance composed of
two conserved "charges", like the baryon number and the isospin
charge in heavy ion collisions at finite baryon density, the ratio
between the two charges is fixed only as long as the system
remains in one of the two pure phases. In the mixed phase, the
concentration in each of the regions of one phase or the other may
be different. Their values are restricted only by the conservation
on the total charge numbers. The essential point is that
conservation laws in chemical thermodynamics are global, not
local. The main result of this formalism is that, different from
the so-called Maxwell construction, the pressure in the mixed
phase is not constant and therefore the nuclear incompressibility,
for example, does not vanish \cite{prl}.

Furthermore, the scenario we are going to explore corresponds to
the situation realized in heavy ion collisions experiments at not
too high energy where finite temperature and high compressed
baryon density is reached. In this condition, a not large fraction
of strangeness can be produced and, therefore, we will limit
ourselves to study the deconfinement transition from hadronic
matter into up and down quark matter \cite{fuchs,ditoro2,ditoro}.
We aspect that, in the range of temperature and density
considered, the presence of strange particles does not
significantly affect the main conclusions regarding the relevance
of nonextensive statistical effects on the nuclear EOS.

The paper is organized as follows. In Section \ref{statistics}, we
introduce the basic formalism of the nonextensive statistics. In
Section \ref{hadron}, we study the nonextensive hadronic EOS for
symmetric and asymmetric nuclear matter and we explore the
behavior of meson fields in presence of small deviations from the
standard statistics. In Section \ref{qgp}, we investigate
nonextensive proprieties of the quark-gluon EOS. In Section
\ref{mp}, we study the hadron to quark-gluon phase transition and
the consequent formation of a mixed phase, mainly focusing our
study in the variation of the first critical transition density
for various set of parameters. Finally, we summarize our
conclusions in Section \ref{conclusion}.

\section{Basic assumptions in nonextensive statistics}\label{statistics}

Nonextensive statistical mechanics introduced by Tsallis is a
generalization of the common Boltzmann-Gibbs statistical mechanics
\cite{tsallis,GMTsallis,book2}. It is based upon the introduction of
the following entropy
\begin{eqnarray}
S_q[f]=\frac{1}{q-1}\, \left(1-\int[f({\bf x})]^q
\,d\Omega\right)\; ,\ \ \ \left(\int f({\bf x})\,d\Omega=1\right)
\, , \label{eq:GMTsallis}
\end{eqnarray}
where $f({\bf x})$ stands for a normalized probability distribution,
${\bf x}$ and $d\Omega$ denoting, respectively, a generic point and
the volume element in the corresponding phase space. Here and in the
following we set the Boltzmann and the Planck constant equal to
unity. The real parameter $q$ determines the degree of
non-additivity exhibited by the entropy form (\ref{eq:GMTsallis}).

The generalized entropy has the usual properties of positivity,
equiprobability, concavity and irreversibility, preserves the
whole mathematical structure of thermodynamics (Legendre
transformations). In the limit $q\rightarrow 1$, the entropic form
(\ref{eq:GMTsallis}) becomes additive and reduces to the standard
Boltzmann-Gibbs entropy
\begin{equation}
S_1=-\int f({\bf x})\, \ln f({\bf x})\, d\Omega\, .
\end{equation}

Peculiarity of the Tsallis generalized thermostatistics is that if
we have two statistically independent subsystems $A$ e $B$,
described, respectively, by the individual probability density
$f^{(A)}$ and $f^{(B)}$ and we call $f^{(A+B)}({\bf x}_A,{\bf
x}_B)=f^{(A)}({\bf x}_A)\,f^{(B)}({\bf x}_B)$ the joint
probability density of a composite system $A+B$, the nonadditive
(nonextensive) character of $S_q$ is summarized in the relation
\cite{book2}
\begin{eqnarray}
S_q[f^{(A+B)}]=S_q[f^{(A)}]+S_q[f^{(B)}]) +(1-q)
S_q[f^{(A)}]\,S_q[f^{(B)}] \, . \label{next}
\end{eqnarray}
In the limit $q\rightarrow 1$, the third term in right hand side
of Eq.(\ref{next}) vanishes and the above equation reduces to the
well-known additivity (extensivity) relation of the
Boltzmann-Gibbs logarithmic entropy. Here, the word nonextensive
should be associated with the fact that the total energy of
long-range-interacting mechanical systems is nonextensive, in
contrast with the case of short-range-interacting systems, whose
total energy is extensive in the thermodynamical sense
\cite{book2}.

Second crucial assumption on nonextensive statistics is the
introduction of the $q$-mean value (or escort mean value) of a
physical observable $A({\bf x})$
\begin{equation}
 \displaystyle\langle A\rangle_q=\frac{\int A({\bf x})\,[f({\bf x})]^q
d\Omega} {\int [f({\bf x})]^q d\Omega} \, . \label{escort}
\end{equation}

The probability distribution can be obtained maximizing the
measure $S_q$ under appropriate constraints related to the
previous definition of the $q$-mean value. In this context, it is
important to observe that the Tsallis classical distribution can
be seen as a superposition of Boltzmann distributions with
different temperatures which have a mean value corresponding to
the temperature appearing in the Tsallis distribution. The
nonextensive $q$ parameter is related to the fluctuation in the
temperature and describes the spread around the average value of
the Boltzmann temperature \cite{wilk1}.

Following the above prescriptions, it is possible to obtain the
associate quantum mean occupation number of particles species $i$
in a grand canonical ensemble. For a dilute gas of particles and
for small deviations from the standard statistics ($q\approx 1$,)
it can be written as \cite{buyu,silva2}
\begin{equation}
n_i=\frac{1} { [1+(q-1)\, \beta(E_i-\mu_i)]^{1/(q-1)} \pm 1} \, ,
\label{eq:distribuzione}
\end{equation}
where $\beta=1/T$ and the sign $(+1)$ is for fermions and $(-1)$
for bosons. Naturally, for $q\rightarrow1$ the above quantum
distribution reduces to the standard Fermi-Dirac and Bose-Einstein
distribution. Let us observe that nonextensive statistical effects
vanishes approaching to zero temperature. This is the reason for
which nonextensive effects could be significantly relevant in high
energy heavy ion collision and probably, in the protoneutron star.
In addition, in high density quark-gluon matter the color magnetic
field remains unscreened (in leading order) and long-range color
magnetic interaction should be present at any finite temperature,
thus QGP appears to be an ideal candidate for finding some
nonextensive behavior.

Finally, let us observe that when the entropic $q$ parameter is
smaller than one, the above distribution have a natural high
energy cut-off which implies that the energy tail is depleted;
when $q$ is greater than one, the cut-off is absent and the energy
tail of the particle distribution (for fermions and bosons) is
enhanced. Hence the nonextensive statistics entails a sensible
difference of the power-law particle distribution shape in the
high energy region with respect to the standard statistics. In
this context, it is relevant to observe that in Ref.
\cite{wilknjl}, the authors postulate a modified quantum
distribution function for fermions and bosons at the scope of
satisfy the particle-hole symmetry, both for $q>1$ and $q<1$. In
the present work we will focus our study for small deviations from
the standard statistics and for values of $q>1$, because these
values were obtained in several phenomenological studies in high
energy heavy ion collisions (see, for example, Ref.s
\cite{albe,wilk2,physicaA2008,cley,chinellato}). We have
explicitly verified that, for the values $q>1$ considered in this
investigation, the prescription introduced in Ref. \cite{wilknjl}
does not affect the results that we are going to obtain and,
therefore, we adopt the original formulation of nonextensive
statistics. Furthermore, it is proper to remember that in a
relativistic mean field theory, considered in this investigation,
baryons are assumed as Dirac quasiparticles moving in classical
meson fields, the field operator are replaced by their expectation
values and the contributions coming from the Dirac sea are
neglected.

\section{Nonextensive hadronic equation of state}
\label{hadron}

In this Section we study the nonextensive hadronic EOS in the
framework of a relativistic mean field theory in which nucleons
interact through the nuclear force mediated by the exchange of
virtual isoscalar-scalar ($\sigma$), isoscalar-vector ($\omega$)
and isovector-vector ($\rho$) meson fields
\cite{walecka,boguta,glen}. As quoted in the Introduction, a
similar approach has been studied in Ref. \cite{silva} for pure
neutron matter and for symmetric nuclear matter (thus, without
considering the effects of the $\rho$ meson field) focusing
principally the attention to a different range of density and
temperature considered in this paper. Here, we are going to study
the hadronic EOS at the scope of emphasize several features
previously not investigated that result to be crucial for our
following purposes.

The Lagrangian density describing hadronic matter can be written
as
\begin{eqnarray}
{\cal L}={\cal L}_{QHD}+{\cal L}_{\rm qfm} \, ,\label{totl}
\end{eqnarray}
where \cite{glen}
\begin{eqnarray}\label{eq:1}
\!\!\!\!\!\!\!\!\!\!\!\!\!\!\!{\cal L}_{QHD}&=&
\bar{\psi}[i\gamma_{\mu}\partial^{\mu}-(M- g_{\sigma}\sigma)
-g{_\omega}\gamma_\mu\omega^{\mu}-g_\rho\gamma^{\mu}\vec\tau\cdot
\vec{\rho}_{\mu}]\psi
+\frac{1}{2}(\partial_{\mu}\sigma\partial^{\mu}\sigma-m_{\sigma}^2\sigma^2)
\nonumber\\
\!\!\!\!\!\!\!\!\!\!\!\!\!\!\!&&-U(\sigma)+\frac{1}{2}m^2_{\omega}\omega_{\mu}
\omega^{\mu}
+\frac{1}{2}m^2_{\rho}\vec{\rho}_{\mu}\cdot\vec{\rho}^{\;\mu}
-\frac{1}{4}F_{\mu\nu}F^{\mu\nu}
-\frac{1}{4}\vec{G}_{\mu\nu}\vec{G}^{\mu\nu}\,,
\end{eqnarray}
and $M=939$ MeV is the vacuum baryon mass. The field strength
tensors for the vector mesons are given by the usual expressions
$F_{\mu\nu}\equiv\partial_{\mu}\omega_{\nu}-\partial_{\nu}\omega_{\mu}$,
$\vec{G}_{\mu\nu}\equiv\partial_{\mu}\vec{\rho}_{\nu}-\partial_{\nu}\vec{\rho}_{\mu}$,
and $U(\sigma)$ is a nonlinear potential of $\sigma$ meson
\begin{eqnarray}
U(\sigma)=\frac{1}{3}a\sigma^{3}+\frac{1}{4}b\sigma^{4}\,,
\end{eqnarray}
usually introduced to achieve a reasonable compression modulus for
equilibrium nuclear matter.

Following Ref.s \cite{muller_npa,lava_prc}, ${\cal L}_{\rm qfm}$
in Eq.(\ref{totl}) is related to a (quasi) free gas of pions with
an effective chemical potential (see below for details).
%, especially
%important in regime of low density and high temperature.

The field equations in a mean field approximation are
\begin{eqnarray}
&&(i\gamma_{\mu}\partial^{\mu}-(M- g_{\sigma}\sigma)-
g_\omega\gamma^{0}\omega-g_\rho\gamma^{0}{\tau_3}\rho)\psi=0\,, \\
&&m_{\sigma}^2\sigma+ a{{\sigma}^2}+ b{{\sigma}^3}=
g_\sigma<\bar\psi\psi>=g_\sigma{\rho}_S\,, \\
&&m^2_{\omega}\omega=g_\omega<\bar\psi{\gamma^0}\psi>=g_\omega\rho_B\,,\\
&&m^2_{\rho}\rho=g_\rho<\bar\psi{\gamma^0}\tau_3\psi>=g_\rho\rho_I\,,
\label{eq:MFT}
\end{eqnarray}
where $\sigma=\langle\sigma\rangle$,
$\omega=\langle\omega^0\rangle$ and $\rho=\langle\rho^0_3\rangle$
are the nonvanishing expectation values of meson fields, $\rho_I$
is the total isospin density, $\rho_B$ and $\rho_S$ are the baryon
density and the baryon scalar density, respectively. They are
given by
\begin{eqnarray}
&&\rho_{B}=2 \sum_{i=n,p} \int\frac{{\rm
d}^3k}{(2\pi)^3}[n_i(k)-\overline{n}_i(k)]\,, \label{eq:rhob} \\
&&\rho_S=2 \sum_{i=n,p} \int\frac{{\rm
d}^3k}{(2\pi)^3}\,\frac{M_i^*}{E_i^*}\,
[n_i^q(k)+\overline{n}_i^{\,q}(k)]\,, \label{eq:rhos}
\end{eqnarray}
where $n_i(k)$ and $\overline{n}_i(k)$ are the $q$-deformed
fermion particle and antiparticle distributions given in
Eq.(\ref{eq:distribuzione}); more explicitly, in this context, we
have
\begin{eqnarray}
n_i(k)=\frac{1} { [1+(q-1)\,\beta(E_i^*(k)-\mu_i^*)
]^{1/(q-1)} + 1} \label{eq:distribuz} \, , \\
\overline{n}_i(k)=\frac{1} {[1+(q-1)\,\beta(E_i^*(k)+\mu_i^*)
]^{1/(q-1)} + 1} \, . \label{eq:distribuz2}
\end{eqnarray}

The nucleon effective energy is defined as
${E_i}^*(k)=\sqrt{k^2+{{M_i}^*}^2}$, where ${M_i}^*=M_{i}-g_\sigma
\sigma$. The effective chemical potentials $\mu_i^*$  are given in
terms of the meson fields as follows
\begin{eqnarray}
\mu_i^*={\mu_i}-g_\omega\omega -\tau_{3i} g_{\rho}\rho \, ,
\label{mueff}
\end{eqnarray}
where $\mu_i$ are the thermodynamical chemical potentials
$\mu_i=\partial\epsilon/\partial\rho_i$. At zero temperature they
reduce to the Fermi energies $E_{Fi} \equiv
\sqrt{k_{Fi}^2+{M_i^*}^2}$ and the nonextensive statistical effects
disappear. The meson fields are obtained as a solution of the field
equations in mean field approximation and the related meson-nucleon
couplings ($g_\sigma$, $g_\omega$ and $g_\rho$) are the free
parameters of the model. In the following, they will be fixed to the
parameters set marked as GM2 of Ref.\cite{glen}.

The thermodynamical quantities can be obtained from the
thermodynamic po\-tential in the standard way. More explicitly,
the baryon pressure $P_B$ and the energy density $\epsilon_B$ can
be written as
\begin{eqnarray}
\hspace{-2cm}&&P_B=\frac{2}{3} \sum_{i=n,p} \int \frac{{\rm
d}^3k}{(2\pi)^3} \frac{k^2}{E_{i}^*(k)}
[n_i^q(k)+\overline{n}_i^q(k)] -\frac{1}{2}m_\sigma^2\sigma^2 -
U(\sigma)+
\frac{1}{2}m_\omega^2\omega^2+\frac{1}{2}m_{\rho}^2 \rho^2\,,\label{eq:eos}\\
\hspace{-2cm}&&\epsilon_B= 2 \sum_{i=n,p}\int \frac{{\rm
d}^3k}{(2\pi)^3}E_{i}^*(k) [n_i^q(k)+\overline{n}_i^q(k)]
+\frac{1}{2}m_\sigma^2\sigma^2+U(\sigma)
+\frac{1}{2}m_\omega^2\omega^2+\frac{1}{2}m_{\rho}^2 \rho^2\, .
\label{eq:eos2}
\end{eqnarray}

It is important to observe that Eq.s(\ref{eq:rhos}),
(\ref{eq:eos}) and (\ref{eq:eos2}) apply to $n_i^q\equiv(n_i)^q$
rather than $n_i$ itself, this is a direct consequence of the
basic prescription related to the $q$-mean expectation value in
nonextensive statistics \cite{book2,pla2002} (this recipe was not
adopted in Ref.\cite{silva}). In addition, since all equations
must be solved in a self-consistent way, the presence of
nonextensive statistical effects influences the many-body
interaction mediated by the meson fields.

Especially in regime of low density and high temperature the
contribution of the lightest mesons to the thermodynamical
potential (and, consequently, to the other thermodynamical
quantities) becomes relevant. As quoted before, following Ref.s
\cite{muller_npa,lava_prc}, we have included the contribution of
pions considering them as a (quasi) ideal gas of nonextensive
bosons with effective chemical potentials expressed in terms of
the corresponding effective baryon chemical potentials. More
explicitly, for $\pi^+$ mesons we have
$\mu_{\pi^+}=\mu_C\equiv\mu_p-\mu_n$, where $\mu_C$ is the
electric charge chemical potential. Thus, the corresponding
effective pion chemical potential can be written as
\begin{eqnarray}
\mu_{\pi^+}^*\equiv\mu_p^*-\mu_n^*= \mu_p-\mu_n-2\,g_\rho\,\rho\,
, \label{mueff_m1}
\end{eqnarray}
where the last equivalence follows from Eq.(\ref{mueff}).
Therefore, the $\rho$ meson field couples to the total isospin
density, which receives a contribution from nucleons and pions.

%%%%%%%%%%%%%%%%%%%%%%%%%%%%%%%%%%%%%%%%%%%%%%%%%%%%%%%%%%%%%%%%%%%%%%%%%%%%%

%In the following, we will concentrate our study in the range of
%baryon chemical potential from $200$ MeV to $1.2$ GeV, that
%corresponds approximately to a baryon density from
%$\rho_{B}\approx 0.1\,\rho_0$ to $\rho_B\approx\,4\rho_{0}$ (where
%$\rho_0$ is the nuclear saturation density) with values of
%temperature in the range of $60\div 120$ MeV. This values can be
%reached in high energy heavy ion collisions at high baryon
%density.

Let us start our numerical investigation by considering the
behavior of $\sigma$, $\omega$ and $\rho$ meson fields at a fixed
value $Z/A=0.4$, for different values of temperature and
nonextensive parameter $q$. Because meson fields have their source
in the baryon and scalar density, which are very sensible to the
behavior of the mean occupation number, all meson fields appear to
be significantly changed in presence of nonextensive effects.

\begin{figure}
\begin{center}
\resizebox{0.6\textwidth}{!}{%
\includegraphics{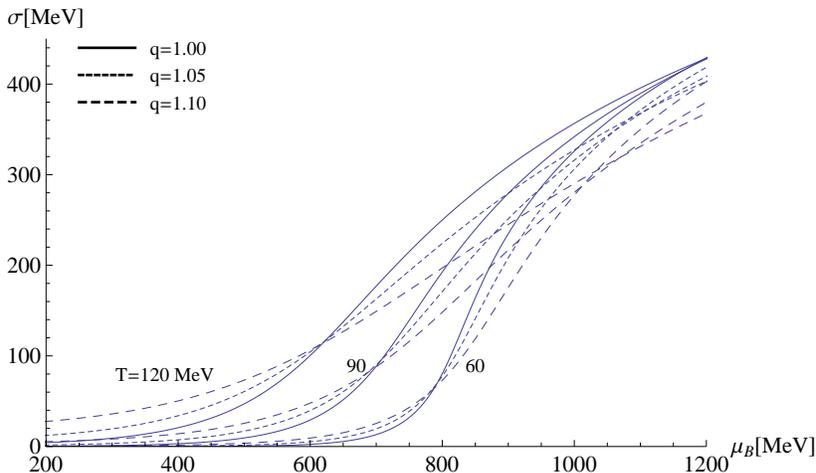}} \caption{The $\sigma$ meson field as a
function of baryon chemical potential for different values of
temperature (in units of MeV) and $q$.} \label{fig:sigma}
\end{center}
\end{figure}

In Fig. \ref{fig:sigma}, we show the $\sigma$ meson field as a
function of the baryon chemical potential $\mu_B$. It is
interesting to observe that at lower $\mu_B$, in presence of
nonextensive effects, the value of the meson field is
significantly increased for all values of temperature respect to
the standard case, the other way round happens at higher $\mu_B$.
This important feature is due to the fact that, as already
remarked in Section \ref{statistics}, for $q>1$ and fixed baryon
density (or $\mu_B$), the (normalized) mean occupation function is
enhanced at high values of its argument and depressed at low
values. Being the argument of the mean occupation function
$x_i=\beta (E_i^*-\mu_i^*)$, in the integration over momentum
(energy), at lower $\mu_{B}$ (corresponding to lower values of the
effective particle chemical potential $\mu_i^*$) the enhanced
Tsallis high energy tail weighs much more that at higher $\mu_{B}$
where depressed low energy effects prevail and the mean occupation
number results to be bigger for the standard Fermi-Dirac
statistics. Concerning the antiparticle contribution, the argument
of $\overline{n}_i$ is $\overline{x}_i=\beta (E_i^*+\mu_i^*)$ and
the Tsallis enhancement at high energy tail is favored also at
higher $\mu_B$. At the same time, higher temperatures (where
antiparticle contribution are more relevant) reduce the value of
the argument of $n_i$ and $\overline{n}_i$, favoring the extensive
distribution. These effects are much more evident for the scalar
density $\rho_S$ (self-consistently related to the $\sigma$ meson
field) where appears $(n_i)^q$ and particle and antiparticle
contributions are summed.
%and take place the ratio $M^*_i/E^*_i$.
%The shape of the $\sigma$ meson field remains almost the same at
%the different temperatures, but anticipates the values of $\mu_B$
%where standard statistics becomes dominant. This is due to the
%fact that, at fixed values $\mu_{B}$, higher temperatures reduce
%the value of the argument $x_i$ of the $n_i$, favoring the
%extensive distribution.
The same effect involves also the nucleon effective mass
$M^*=M-g_{\sigma}\sigma$, which becomes, respect to the standard
case, smaller for lower values of $\mu_{B}$ and bigger for higher
values, with very relevant consequences for the hadronic EOS
\footnote{In Ref. \cite{silva}, the nucleon effective mass as a
function of temperature always diminishes respect to standard
statistics, this behavior is a consequence of the fact that it is
plotted only at $\rho_B=0$.}.

\begin{figure}
\begin{center}
\resizebox{0.6\textwidth}{!}{%
\includegraphics{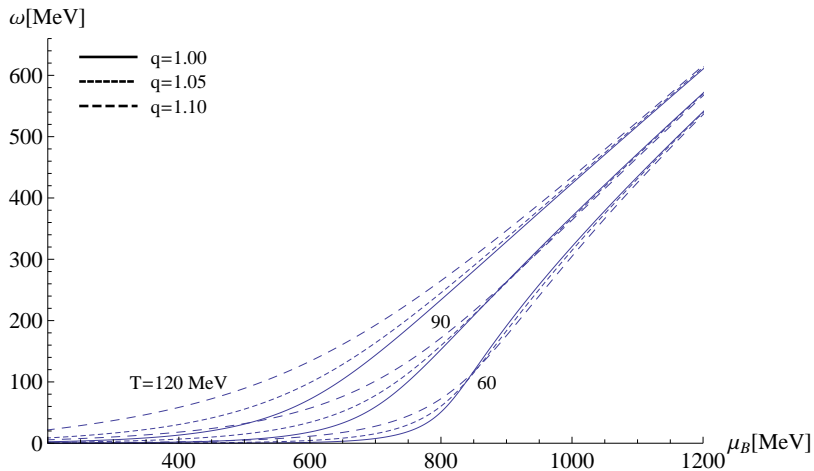}
} \caption{The $\omega$ meson field as a function of the baryon
chemical potential for different values of temperature and $q$.}
\label{fig:omega} \end{center}
\end{figure}

In Fig. \ref{fig:omega}, we report the $\omega$ meson field as a
function of the baryon chemical potential for different values of
temperature and $q$. In this case the situation is different from
the $\sigma$ meson, because the $\omega$ field have its source in
the baryon density $\rho_B$ where appears $n_i$ and particle and
antiparticle contributions are subtracted. At lower temperatures
($T=60$ MeV), antiparticle contributions are negligible and we
have a behavior similar (although less evident) to the $\sigma$
field. At higher temperatures ($T=120$ MeV), the contributions of
antiparticle increase and nonextensive effects vanish at higher
$\mu_B$.
%Therefore, concerning the $\omega$ meson field, the
%relevance of nonextensive effects to the EOS results to be
%appreciable only around and below $2\div 3 \, \rho_0$.

\begin{figure}
\begin{center}
\resizebox{0.6\textwidth}{!}{%
\includegraphics{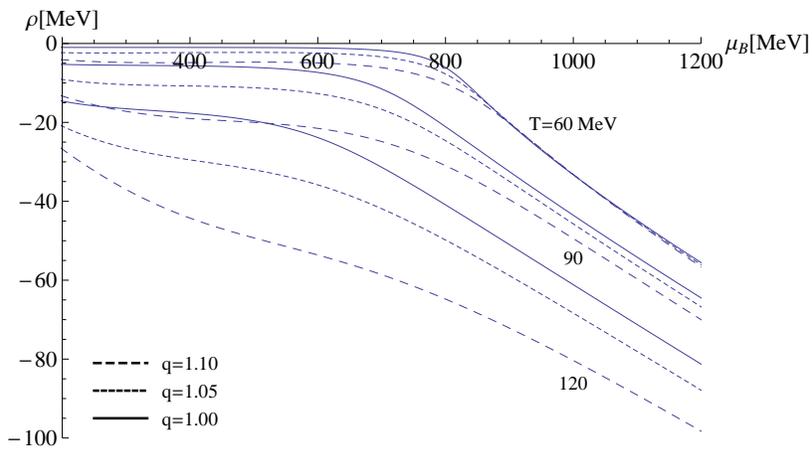}
} \caption{The $\rho$ meson field as a function of baryon chemical
potential for different values of temperature and $q$.}
\label{fig:rho} \end{center}
\end{figure}

Finally, in Fig. \ref{fig:rho}, we report the behavior of the
$\rho$ meson field which depends from the isospin density (let us
remember that we have fixed $Z/A=0.4$). Similar arguments as done
for the $\omega$ meson applies also in this case. The valuable
increasing of its absolute value, also for weakly asymmetric
nuclear matter, makes $\rho$ meson very relevant in the hadronic
EOS, especially at not too large $\mu_{B}$.

\begin{figure}
\begin{center}
\resizebox{1.0\textwidth}{!}{%
\includegraphics{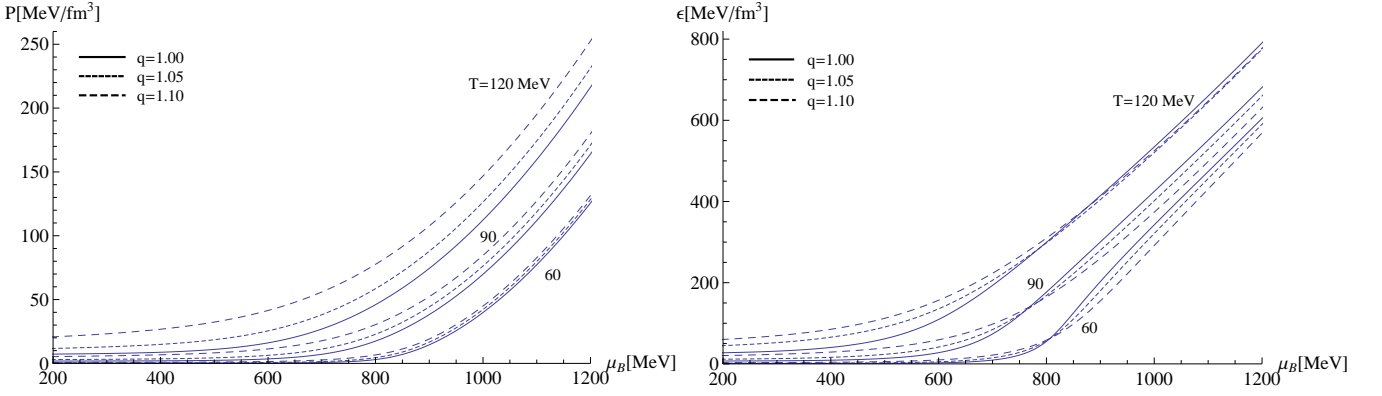}
} \caption{Pressure (left panel) and energy density (right panel)
versus baryon chemical potential for different values of
temperature and $q$.} \label{fig:P} \end{center}
\end{figure}

In Fig. \ref{fig:P}, the total pressure $P$ and energy density
$\epsilon$ are plotted as a function of $\mu_B$ for different
values of temperature and $q$. The different behavior from $P$ and
$\epsilon$ reflects essentially the nonlinear combinations of the
meson fields and the different functions under integration in Eq.s
(\ref{eq:eos}) and (\ref{eq:eos2}). Concerning the pressure, we
have that becomes stiffer by increasing the $q$ parameter. On the
other hand, the behavior of the energy density presents features
very similar to the $\sigma$ field one. At low $\mu_{B}$,
nonextensive effects make the energy density greater with respect
to the standard case. At medium-high $\mu_B$, the standard ($q=1$)
component of the energy density becomes dominant, this effect is
essentially due to the reduction of the $\sigma$ field for $q>1$.
The intersection point depends, naturally, on the physical
parameters of the system.

\section{Nonextensive QGP equation of state}\label{qgp}

Concerning the nonextensive quark-gluon EOS, due to its
simplicity, we adopt the MIT bag model \cite{mit}. In this model,
quark matter is described as a gas of free quarks and all
non-perturbative effects are simulated by the bag constant $B$
which represents the pressure of the vacuum.

%As previously discussed, in this investigation we are limiting our
%study to the two-flavor ($f=u, \, d$) massless quarks. This
%appears rather well justified for the application to heavy ion
%collisions at relativistic (but not ultra-relativistic) energies,
%the fraction of strangeness produced at these energies being small
%\cite{ditoro2,fuchs}.

Following this line, the pressure, energy density and baryon
density for a relativistic Fermi gas of quarks in the framework of
nonextensive statistics can be written, respectively, as
\begin{eqnarray}
&P_q& =\frac{\gamma_f}{3} \sum_{f=u,d} \int^\infty_0 \frac{{\rm
d}^3k}{(2\pi)^3} \,\frac{k^2}{e_f}\,
[n_f^q(k)+\overline{n}_f^q(k)]
%\nonumber \\&&\;\;\;\;
-B\,, \label{bag-pressure}\\
&\epsilon_q& =\gamma_f \sum_{f=u,d}  \int^\infty_0 \frac{{\rm
d}^3k}{(2\pi)^3} \,e_f\, [n_f^q(k)+\overline{n}_f^q(k)]
\label{bag-energy}
%\nonumber \\&&\;\;\;\;
+B\,, \\
&\rho_q& =\frac{\gamma_f}{3} \sum_{f=u,d} \int^\infty_0 \frac{{\rm
d}^3k}{(2\pi)^3}  \,[n_f(k)-\overline{n}_f(k)]\, ,
\label{bag-density}
\end{eqnarray}
where the quark degeneracy for each flavor is $\gamma_f=6$,
$e_f=(k^2+m_f^2)^{1/2}$, $n_f(k)$ and $\overline{n}_f(k)$ are the
$q$-deformed particle and antiparticle quark distributions
\begin{eqnarray}
n_f(k)=\frac{1} { [1+(q-1)(e_f(k)-\mu_f)/T
]^{1/(q-1)} + 1} \, , \\
\overline{n}_f(k)=\frac{1}{[1+(q-1)(e_f(k)+\mu_f)/T ]^{1/(q-1)} +
1} \, .
\end{eqnarray}

Similar expressions for the pressure and the energy density can be
written for gluons treating them as a massless $q$-deformed Bose
gas with zero chemical potential. Explicitly, we can calculate the
nonextensive pressure $P_g$ and energy density $\epsilon_g$ for
gluons as
\begin{eqnarray}
&P_g& =\frac{\gamma_g}{3} \int^\infty_0 \frac{{\rm
d}^3k}{(2\pi)^3}
\,\frac{k}{[1+(q-1)\,k/T]^{q/(q-1)} - 1}\,, \label{gluon-press}\\
&\epsilon_g& =3\, P_g \, , \label{gluon-energy}
\end{eqnarray}
with the gluon degeneracy factor $\gamma_g=16$. In the limit
$q\rightarrow 1$, one recovers the usual analytical expression:
$P_g=8\pi^2/45\,T^4$.

Let us note that, since one has to employ the fermion (boson)
nonexten\-sive distributions, the results are not analytical, even
in the massless quark approximation. Hence a numerical evaluations
of the integrals in Eq.s~(\ref{bag-pressure})--(\ref{bag-density})
and (\ref{gluon-press}) must be performed.

\begin{figure}
\begin{center}
\resizebox{0.6\textwidth}{!}{%
  \includegraphics{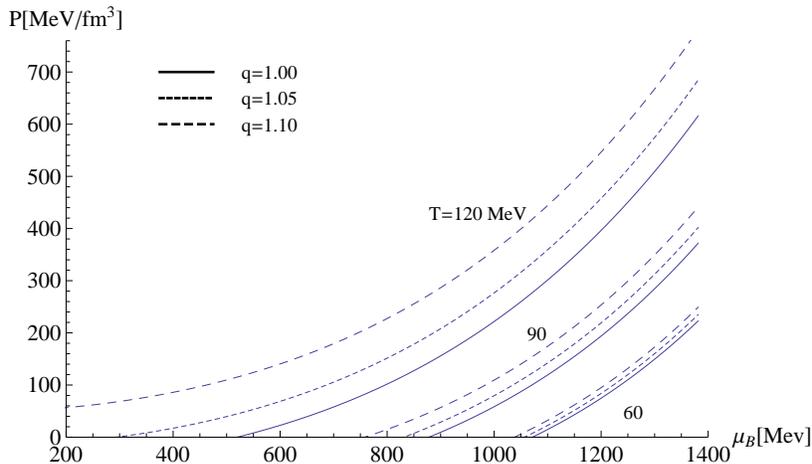}
} \caption{Pressure of the quark-gluon phase as a function of
baryon chemical potential for different values of temperature and
$q$.} \label{fig:PQGP}
\end{center}
\end{figure}

In Fig. \ref{fig:PQGP}, we report the total pressure as a function
of the baryon chemical potential for massless quarks and gluons,
for different values of $q$ and at fixed value of $Z/A=0.4$. The
bag constant is set equal to $B^{1/4}$=190 MeV. In presence of
nonextensive effects, as in the case of hadronic phase, the
pressure is significantly increased even for small deviations from
standard statistics.

\section{The hadron to quark-gluon phase transition}\label{mp}

In this Section we are going to investigate the phase transition
from hadronic matter to QGP at finite temperature and baryon
chemical potential in the framework of nonextensive statistics.

At this scope, we use the Gibbs formalism applied to systems where
more than one conserved charge is present \cite{glenprd}. In fact,
because we are going to describe the nuclear EOS, we have to
require the global conservation of two "charges": baryon number
and electric charge. Each conserved charge has a conjugated
chemical potential and the systems is described by two independent
chemical potentials: $\mu_B$ and $\mu_C$. The structure of the
mixed phase is obtained by imposing the following Gibbs conditions
for chemical potentials and pressure
\begin{eqnarray}
&&\mu_B^{(H)} = \mu_B^{(Q)} \, , \ \ \  \mu_C^{(H)} = \mu_C^{(Q)}
\, , \\
&&P^H (T,\mu_B,\mu_C)=P^Q (T,\mu_B,\mu_C) \, .
\end{eqnarray}
Therefore, at a given baryon density $\rho_B$ and at a given net
electric charge density $\rho_C=Z/A\, \rho_B$, the chemical
potentials $\mu_B$ are $\mu_C$ are univocally determined by the
following equations
%
%\begin{eqnarray}
%&&\rho_B=(1-\chi)\,\sum_{i=p,n} b_i\,\rho_i^H(T,\mu_B,\mu_C)
%+\chi \,\sum_{i=u,d} b_i\,\rho_i^Q(T,\mu_B,\mu_C) \, ,\\
%&&\rho_C=(1-\chi)\,\sum_{i=p,n} c_i\, \rho_i^H(T,\mu_B,\mu_C)
%+\chi \,\sum_{i=u,d} c_i\,\rho_i^Q(T,\mu_B,\mu_C) \, ,
%\end{eqnarray}
\begin{eqnarray}
&&\rho_B=(1-\chi)\,\rho_B^H(T,\mu_B,\mu_C)
+\chi \,\rho_B^Q(T,\mu_B,\mu_C) \, ,\\
&&\rho_C=(1-\chi)\,\rho_C^H(T,\mu_B,\mu_C) +\chi
\,\rho_C^Q(T,\mu_B,\mu_C) \, ,
\end{eqnarray}
where $\rho_B^{H(Q)}$ and $\rho_C^{H(Q)}$ are, respectively, the
net baryon and electric charge densities in the hadronic (H) and
in the quark (Q) phase and $\chi$ is the fraction volume of
quark-gluon matter in the mixed phase. In this way we can find out
the phase coexistence region, for example, in the $(T,\mu_B)$
plane. We are particularly interested in the lower baryon density
(baryon chemical potential) border, i.e. the first critical
transition density $\rho_{\rm cr}^I$ ($\mu_{\rm cr}^I$), in order
to check the possibility of reaching such conditions in a
transient state during a heavy-ion collision at relativistic
energies.

\begin{figure}
\begin{center}
\resizebox{0.6\textwidth}{!}{%
  \includegraphics{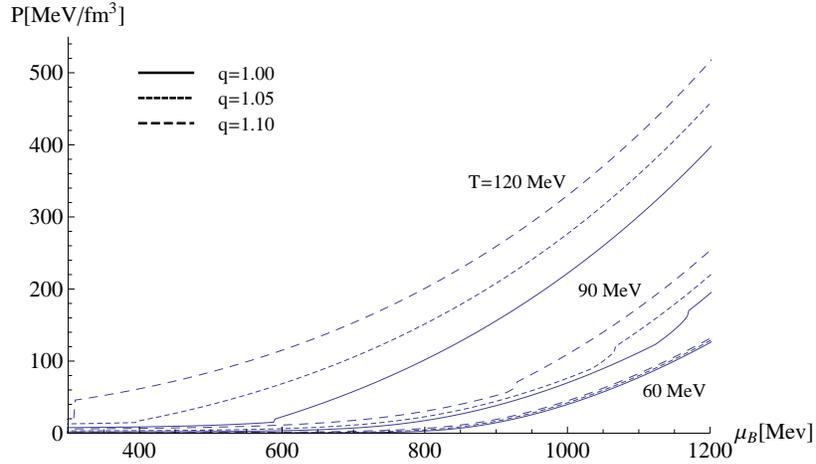} } \caption{Pressure versus baryon chemical potential in
the mixed phase for different values of temperature and $q$.}
\label{fig:PMP} \end{center}
\end{figure}

In Fig. \ref{fig:PMP}, we report the pressure as a function of
baryon chemical potential; as before, we have set $Z/A=0.4$ and
$B^{1/4}=190$ MeV.
\begin{figure}
\begin{center}
\resizebox{1.0\textwidth}{!}{%
  \includegraphics{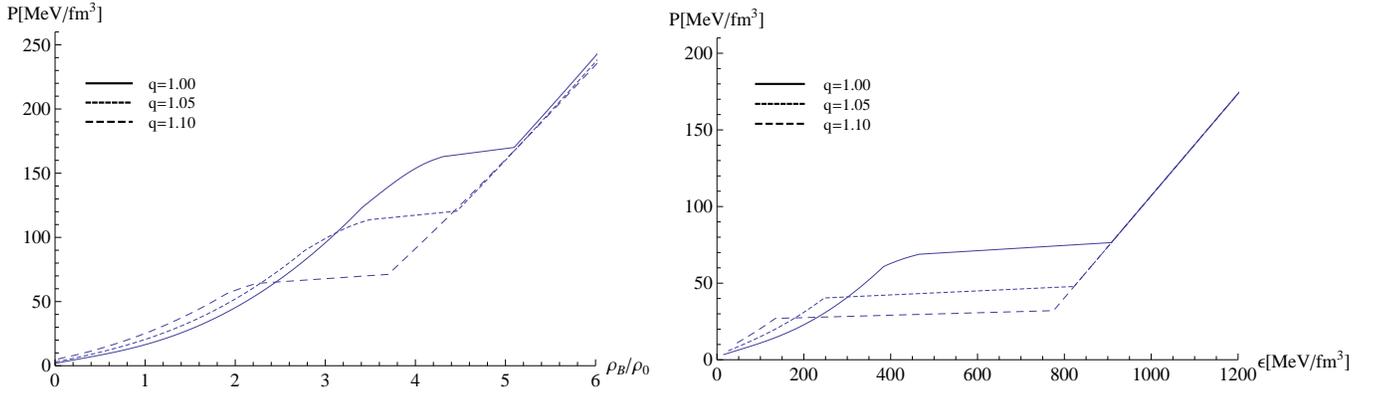} } \caption{Pressure as a function of baryon density
  (left panel) and energy density (right panel) in the mixed
  phase for different values of $q$. The temperature is fixed at
$T=90$ MeV.} \label{fig:P-rhob-e}
\end{center}
\end{figure}
We can see that at $T=60$ MeV, in the range of the considered
$\mu_B$, the system is in a pure hadronic phase even for the
nonextensive index $q=1.1$. At $T=90$ and $120$ MeV, we have both
the first and the second transition critical density for the
considered values of $q$. In presence of nonextensive effects, the
values of the critical densities result to be sensibly reduced
with respect to the standard case. This matter of fact is more
evident in Fig. \ref{fig:P-rhob-e}, where we report the pressure
at $T=90$ MeV as a function of baryon density (in units of nuclear
saturation density $\rho_{0}=0.153$ fm$^{-3}$) (left panel) and
energy density (right panel). It is interesting to observe that
pressure as a function of baryon density (or energy density) is
stiffer in the pure hadronic phase for $q>1$ but appears a strong
softening in the mixed phase. This feature results in significant
changes in the incompressibility and may be particularly important
in identifying the presence of nonextensive effects in high energy
heavy ion collisions experiments. Related to this aspect, let us
observe that possible indirect indications of a significative
softening of the EOS at the energies reached at AGS have been
discussed several times in the literature
\cite{iva,sahu,stocker,isse,prl}.

In Table \ref{tab:parametricritici}, we report the critical baryon
densities and baryon chemical potentials at the beginning (index
$I$) and at the end of the mixed phase (index $II$) for different
values of temperature and $q$.

\begin{table}
\begin{center}
\begin{tabular}{ccccc}
\hline\hline
$T=60$ MeV       & $\rho^{I}_{\rm cr}/\rho_0$     & $\rho^{II}_{\rm cr}/\rho_0$ & $ \mu^{I}_{\rm cr}$ [MeV]    &$\mu^{II}_{\rm cr}$ [MeV] \\
\hline
$q=1.00$       & $5.75$              & $9.10$         & $1503$               &$1569$   \\
$q=1.05$       & $5.56$              & $8.88$         & $1472$               &$1537$   \\
$q=1.10$       & $5.33$              & $8.65$         & $1437$               &$1502$   \\
\hline \hline
$T=90$ MeV       & $\rho^{I}_{\rm cr}/\rho_0$     & $\rho^{II}_{\rm cr}/\rho_0$ & $ \mu^{I}_{\rm cr}$ [MeV]    &$\mu^{II}_{\rm cr}$ [MeV] \\
\hline
$q=1.00$       & $3.41$              & $5.09$         & $1123$               &$1170$   \\
$q=1.05$       & $2.77$              & $4.46$         & $1034$               &$1068$   \\
$q=1.10$       & $1.91$              & $3.69$         & $916$                &$927$    \\
\hline \hline
$T=120$ MeV       & $\rho^{I}_{\rm cr}/\rho_0$     & $\rho^{II}_{\rm cr}/\rho_0$ & $ \mu^{I}_{\rm cr}$ [MeV]    &$\mu^{II}_{\rm cr}$ [MeV] \\
\hline
$q=1.00$       & $0.45$              & $1.93$         & $588$               &$616$   \\
$q=1.05$       & $0.20$              & $1.33$         & $383$               &$396$  \\
$q=1.10$       & $0.08$              & $0.71$         & $184$               &$201$   \\
\hline
\end{tabular}
\end{center}
\caption[Critical baryon densities and baryon chemical potentials
at different temperatures]{Critical baryon densities and baryon
chemical potentials at the beginning (index $I$) and at the end
(index $II$) of the mixed phase for different values of
temperature and nonextensive parameter $q$.}
\label{tab:parametricritici}
\end{table}

\begin{figure}
\begin{center}
\resizebox{0.6\textwidth}{!}{%
  \includegraphics{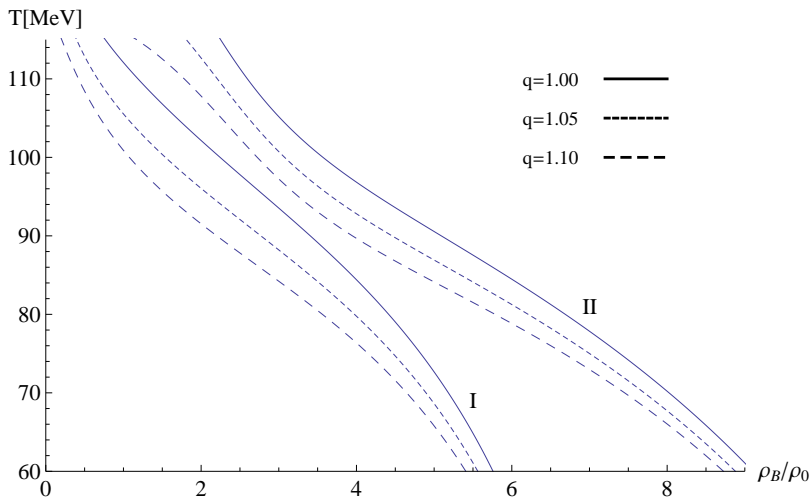} }
  \caption{Phase diagram $T-\rho_B$ for different values of $q$. The
curves with index $I$ and $II$ indicate, respectively, the
beginning and the end of the mixed phase.} \label{fig:RhoCT}
\end{center}
\end{figure}

In Fig. \ref{fig:RhoCT}, it is reported the phase diagram in the
plane $T-\rho_B$ for different values of $q$. The curves labelled
with the index $I$ and $II$ represent, respectively, the beginning
and the end of the mixed phase. For $q>1$, both the first and the
second critical densities are sensibly reduced, even if the shape
of the mixed phase is approximately the same. Related to this
aspect, let us mention that the simplest version of the MIT bag
model, considered in this investigation, appears to be not fully
appropriate to describe a large range of temperature and density.
To overcome this shortcoming, a phenomenological approach can
therefore be based on a density or temperature dependent bag
constant \cite{prl,burgio,rafe2,prasad}. Moreover, as discussed in
the Introduction, in regime of high temperature and small baryon
chemical potential the first order phase transition may end in a
(second order) critical endpoint with a smooth crossover. These
features cannot be incorporated in the considered mean field
approach. In our investigation, because we are focusing to
nonextensive statistical effects on the nuclear EOS, instead of
introducing additional parameterizations, we work with a fixed bag
constant and limit our analysis to a restricted range of
temperature and density, region of particular interest for high
energy compressed nuclear matter experiments.

\begin{figure}
\begin{center}
\resizebox{0.6\textwidth}{!}{%
  \includegraphics{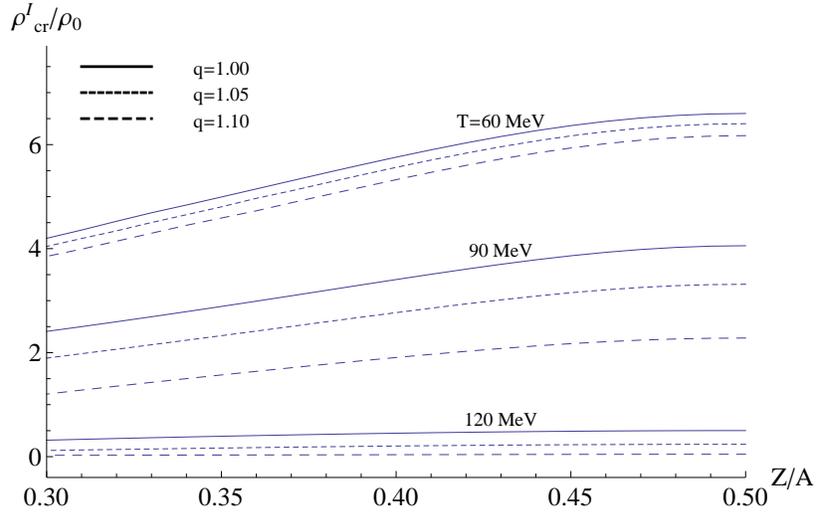} }
  \caption{Variation of the first transition baryon
  density as a function of the net electric charge fraction $Z/A$ and
  for different values of $q$.} \label{fig:RhoCZA}
\end{center}
\end{figure}

Let us now explore in more details the variation of the first
transition baryon density $\rho^{I}_{\rm cr}$ as a function of
different physical parameters. In Fig. \ref{fig:RhoCZA}, we report
the dependence of $\rho^{I}_{\rm cr}$ as a function of $Z/A$ for
different values of $q$. It is interesting to note a significant
reduction of $\rho^{I}_{\rm cr}$ in presence of nonextensive
statistics; this effect is remarkable especially for $T=90$ MeV
and is essentially a consequence of the $\rho$ meson field
behavior in the hadronic phase.

\begin{figure}
\begin{center}
\resizebox{1.0\textwidth}{!}{%
  \includegraphics{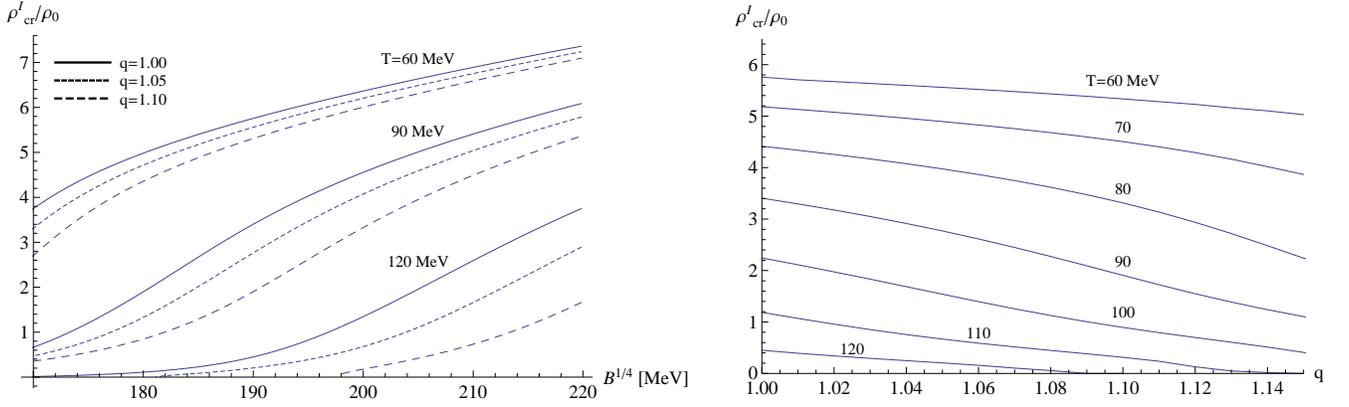} } \caption{Variation of the first
  transition baryon density as a function of the bag constant (left panel)
  and nonextensive index $q$ (right panel) for different temperatures.}
  \label{fig:RhoCBag}
 \end{center}
\end{figure}

In Fig. \ref{fig:RhoCBag} (left panel), we show the first critical
baryon density as a function of the bag constant for different
values of nonextensive parameter $q$. Obviously, by increasing the
bag constant we have a corresponding increasing of $\rho^I_{\rm
cr}$. However, this effect depends on the temperature and
nonextensive parameter $q$.
Finally, in the right panel of Fig. \ref{fig:RhoCBag}, we show the
variation of $\rho^{I}_{\rm cr}$ as a function of the nonextensive
index $q$ for different values of temperature and $B^{1/4}=190$
MeV. At $T=60$ MeV, we can see only a little reduction in the
first critical density also for large deviations from the standard
statistics; on the other hand, the reduction becomes more
pronounced at larger temperatures.
%This
%matter of fact emphasizes the relevance of considering the
%nonextensive statistical effects especially in high energy heavy
%ion collisions where a state of finite temperature and baryon
%density can take place \cite{senger,arsene,bravina}.

\section{Conclusions}\label{conclusion}

To summarize, we have studied the main features of the nuclear EOS
in the hadronic and quark-gluon phase and the possible formation
of a consequent mixed phase in presence of nonextensive
statistical effects.  We have focused our investigation in regime
of finite temperature and baryon chemical potential, reachable in
high-energy heavy-ion collisions, for which the deconfinement
phase transition can be still considered of the first order. From
a phenomenological point of view, the nonextensive index $q$ is
considered here as a free parameter, even if, actually should not
be treated as such because, in principle, it should depend on the
physical conditions generated in the reaction, on the fluctuation
of the temperature and be related to microscopic quantities (such
as, for example, the mean interparticle interaction length, the
screening length and the collision frequency into the parton
plasma). We have restricted our investigation for small deviations
from the standard statistics and for values $q>1$ because, as
quoted in the Introduction, these values were obtained in several
phenomenological studies in high energy heavy ion collisions. In
this context, it is relevant to observe that by fitting
experimental observable at $q>1$, the temperature (or slope)
parameter $T$ is usually minor of the one obtained in the standard
Boltzmann-Gibbs statistics ($q=1$) \cite{albe,wilk2}. This feature
is also present in the considered nuclear equation of state
because, at fixed energy per particle $E/N$, we obtain for $q>1$
lower values of temperature respect to the standard case.
Moreover, let us remember that, in the diffusional approximation,
a value $q>1$ implies the presence of a superdiffusion among the
constituent particles (the mean square displacement obeys to a
power law behavior $\langle x^2\rangle\propto t^\alpha$, with
$\alpha>1$) \cite{tsamem}.

In the first part of the work, we have investigated the hadronic
equation of state and the role played by the meson fields in the
framework of a relativistic mean field model which contains the
basic prescriptions of nonextensive statistical mechanics. We have
shown that, also in presence of small deviations from standard
Boltzmann-Gibbs statistics, the meson fields and, consequently,
the EOS appear to be sensibly modified. In the second part, we
have analyzed the QGP proprieties using the MIT Bag model and also
in this case the EOS becomes stiffer in presence of nonextensive
effects. Finally, we have studied the proprieties of the phase
transition from hadronic matter to QGP and the formation of a
relative mixed phase by requiring the Gibbs conditions on the
global conservation of baryon number and electric charge fraction.
We have seen that nonextensive effects play a crucial role in the
deconfinement phase transition. Moreover, although pressure as a
function of baryon density is stiffer in the hadronic phase, we
have shown that a strong softening in the mixed phase takes place
in presence of nonextensive statistics. Such a behavior implies an
abruptly variation in the incompressibility and could be
considered as a signal of nonextensive statistical effects in high
energy heavy ion collisions.

\end{document}